\documentclass{article}
\usepackage[latin9]{inputenc}
\usepackage{geometry}
\usepackage{amsmath}
\usepackage{amssymb}
\usepackage[unicode=true,pdfusetitle,
 bookmarks=true,bookmarksnumbered=false,bookmarksopen=false,
 breaklinks=false,pdfborder={0 0 1},backref=section,colorlinks=false]
 {hyperref}

\makeatletter

\providecommand{\tabularnewline}{\\}

\usepackage{float}
\usepackage{graphicx}
\usepackage{grffile}
\usepackage{breakurl}

\providecommand{\tabularnewline}{\\}

\usepackage{fullpage}\usepackage[justification=centering]{caption}\usepackage[all]{hypcap}\usepackage{microtype}

\makeatother

\begin{document}

\title{Gaps in nuclear spectra as traces of seniority changes in systems
of both neutrons and protons}

\author{Larry Zamick\\
\textit{Department of Physics and Astronomy, Rutgers University, Piscataway,
New Jersey 08854}\\
 }

\date{\today}
\maketitle
\begin{abstract}
There has been a great deal of attention to the low lying energy spectum
in a nucleus because of the abundance of experimetal data. Likewise
,perhaps to a lesser extent but still significant the high end for
a given configuration has been examined. Here ,using single j shell
calculations as a guide we examine the middle part of the spectum
resulting form single j shell calculations. Seniority arguments are
used to partially explain the midshell behaviours even though in general
seniority is not a good quantum number for mixed systems of neutrons
and protons.
\end{abstract}

\section{Introduction }

One of the first things the present author learned was that for a
system of neutrons and protons the seniority is not a good quantum
number.This is when the author joined a collaboration with John McCullen
and Ben Bayman on wave functions in the f$_{7/2}$ shell{[}1{]}. Only
for a system of particles of one kind e.g. neutrons in the Ca isotopes
is seniority conserved but once one has mixed systems the neutron-proton
interaction strongly mixes states of different seniority. Nevetheless
in this work it will becontend that some remnants of the seniority
conserving J=0 T=1 pairing interaction survive. 

We have recently performed single j shell studies of a system of 3
protons and one neutron (or holes) e.g. $^{96}$Ag as 3 g$_{9/2}$
proton holes and one g$_{9/2}$ neutron hole.{[}2,3{]} We focused
on the yrast T=1 sttes and came up with a (2j-1) rule, namey that
states with total angular momentum I=2j-1 lay very low in energy sometimes
being the ground state. This value of I corresponds to the middle
of the calculated spectrum. The spectrum of $^{96}$Ag is poorly known
but the rule has been verified experimentally for lighter nuclei such
as $^{44}$Sc and $^{52}$Mn. 

.

Table I Experimental and calculated spectra of odd-odd nuclei

\begin{tabular}{|c|c|c|c|}
\hline 
 & I & EXP & TH\tabularnewline
\hline 
\hline 
$^{44}$Sc & 5 & 1.513 & 1.276\tabularnewline
\hline 
 & 6 & 0.271 & 0.381\tabularnewline
\hline 
 & 7 & 0.968 & 1.272\tabularnewline
\hline 
$^{52}$Mn & 5 & 1.254 & 1.404\tabularnewline
\hline 
 & 6 & 0 & 0\tabularnewline
\hline 
 & 7 & 0.870 & 1.819\tabularnewline
\hline 
$^{96}$Ag & 7 & ? & 0.861\tabularnewline
\hline 
 & 8 & 0 & 0\tabularnewline
\hline 
 & 9 & 0.470 & 0.492\tabularnewline
\hline 
h$_{11/2}$ & 9 & ? & 1.30\tabularnewline
\hline 
Q.Q & 10 & ? & 0.21\tabularnewline
\hline 
 & 11 & ? & 0.85\tabularnewline
\hline 
\end{tabular}

In the present work we extend the study to even-even nuclei such as
$^{44}$Ti and $^{96}$ Cd. Our contention will be that there is a
fairly wide gap that separates the lower part of the spectrum from
the upper part.

\section{Calculations.}

In the following 4 tables (II,III, IV and V) we present the energies
of calculated spectra (MeV) as well as the difference in energies
of adjacent spectra (Diff).

TableII Calculated even I spectra of $^{44}$Ti (a) and $^{52}$Fe
(b)

\begin{tabular}{|c|c|c|c|c|}
\hline 
I & INTa & Diffa. & INTb & Diffb\tabularnewline
\hline 
\hline 
0 & 0 &  & 0 & \tabularnewline
\hline 
2 & 1.1613 & 1.1613 & 1.0392 & 1.0392\tabularnewline
\hline 
4 & 2.7900 & 1.6287 & 2.7737 & 1.7345\tabularnewline
\hline 
6 & 4.0618 & 1.2718 & 4.2631 & 1.4634\tabularnewline
\hline 
8 & 6.0842 & 2.0244 & 6.0191 & 1.7830\tabularnewline
\hline 
10 & 7.3839 & 1.3007 & 7.0903 & 1.0712\tabularnewline
\hline 
12 & 7.7022 & 0.3183 & 6.9671 & -0.1232\tabularnewline
\hline 
\end{tabular}

.

Table III Calculated even I spectrum of $^{96}$Cd --INTd

\begin{tabular}{|c|c|c|}
\hline 
I & E(MeV) & Diff.\tabularnewline
\hline 
\hline 
0 & 0.0000 & \tabularnewline
\hline 
2 & 0.2791 & 0.2971\tabularnewline
\hline 
4 & 0.9434 & 0.6463\tabularnewline
\hline 
6 & 1.8344 & 0.8905\tabularnewline
\hline 
8 & 1.9276 & 0.0923\tabularnewline
\hline 
10 & 3.1649 & 1.2373\tabularnewline
\hline 
12 & 3.9119 & 0.7470\tabularnewline
\hline 
14 & 4.1382 & 0.2263\tabularnewline
\hline 
16 & 3.4830 & -0.6552\tabularnewline
\hline 
\end{tabular}

.

.Table IV Calculated Spectrum of $^{96}$Cd--Qi

\begin{tabular}{|c|c|c|}
\hline 
I & E(MeV) & Diff.\tabularnewline
\hline 
\hline 
0 & 0.000 & \tabularnewline
\hline 
2 & 0.8972 & 0.8972\tabularnewline
\hline 
4 & 2.0105 & 1.1133\tabularnewline
\hline 
6 & 3.0576 & 1.0411\tabularnewline
\hline 
8 & 3.4324 & 0.3748\tabularnewline
\hline 
10 & 5.1134 & 1.6810\tabularnewline
\hline 
12 & 5.6409 & 0.5275\tabularnewline
\hline 
14 & 5.7687 & 0.1278\tabularnewline
\hline 
16 & 5.5531 & -0.2156\tabularnewline
\hline 
\end{tabular}

.Table V Calculated odd I spectrum of $^{96}$Cd --INTd

.%
\begin{tabular}{|c|c|c|}
\hline 
I & E(MeV) & Diff.\tabularnewline
\hline 
\hline 
1 & 4.1160 & --\tabularnewline
\hline 
3 & 4.2220 & 0.1060\tabularnewline
\hline 
5 & 4.3708 & 0.1486\tabularnewline
\hline 
7 & 4.4944 & 0.1236\tabularnewline
\hline 
9 & 4.1256 & -0.3688\tabularnewline
\hline 
11 & 5.5640 & 1.4384\tabularnewline
\hline 
13 & 5.8961 & 0.3311\tabularnewline
\hline 
15 & 6.2787 & 0.3836\tabularnewline
\hline 
\end{tabular}

In a single j shell calculation for 2 protons and 2 neutrons in a
single j shell the maximum angular momentum of the 2 protons is 2j-1
and likewise for the 2 neutrons. Hence for the 4 particle system the
maximum anguar momentum I$_{max}$ is equal to 4j-2 and the middle
angular momentum is (2j-1).

We have recently performed single j shell studies of a system of 3
protons and one neutron (or holes) e.g. $^{96}$Ag as 3 g$_{9/2}$
proton holes and one g$_{9/2}$ neutron hole. We focused on the yrast
T=1 states and came up with a (2j-1) rule, namey that states with
total angular momentum I=2j-1 lay very low in energy sometimes being
the ground state. This value of I corresponds to the middle of the
calculated spectrum. The spectrum of $^{96}$Ag is poorly known but
the rule has been verified experimentally for lighter nuclei such
as $^{44}$Sc and $^{52}$Mn. Results from ref {[}3{]} are shown in
Table I.

In the present work we extend the study to even-even nuclei such as
$^{44}$Ti, $^{52}$Fe$^{96}$Cd. Our contention will be that there
is a fairly wide gap that separates the lower part of the spectrum
from the upper part. The spectra are shown in Tables II for the INTa
($^{44}$Ti )and INTb interactions ($^{52}$Fe) {[}2,4{]} ;Table III
for INTd1,4{]} and TableIV for the Qi interaction{[}5{]} (both for
$^{96}$Cd. All the states considered have isospin T=0. We focus only
on even I in these tables. The lower half of the spectrum consists
of states up to I=6 for the f$_{7/2}$ shell and up to I=8 for the
g$_{9/2}$ shell. The midshell angular momenta are I=6 and I =8 respectively.
These gaps, between I=2j+1 and I=2j=1( 8 to 6 and 10 to 8 respectively)
are larger than the neighboring ones. These effects persist for several
different interactions. Besides the INT d interaction of ref {[}1{]}
and Qi{[}3{]} there is the one of Coraggio et al. {[}6{]}. They all
give qualitatively similar results. 

We show more briefly the calculated odd I spectrum for $^{96}$Cd
with the INTd interaction. We see that the first few levels are spaced
very close to each other but these is a sudden gap between I= 9$^{+}$
and I=11$^{+}$ of 1.4384 MeV.

Let us make a brief digression to the highest energy levels. We note
that in $^{44}$Ti the 12$^{+}$ state is correctly predicted to be
higher in energy than the 10$^{+}$ state but in $^{52}$Fe the 12$^{+}$
is lower . This is due to the fact that the J=7 T=0 two-body input
matrix element in $^{54}$Co is smaller than in $^{42}$Sc. These
results are in agreement with experiment . The consequences are that
the 12$^{+}$ state in $^{52}$Fe has a much longer half- life (45.9s)
than the one in $^{44}$Ti(2.1 ns).

In $^{96}$Cd the calculated I=16$^{+}$ state is lower than I=14$^{+}$
(and also 15$^{+}$).This implies that the 16$^{+}$ state is isomeric.
This is in agreement with the experiment of Nara Singh{[}7{]}.

\section{Experimental results compared with theory}

There is not enough experimental data in the g$_{9/2}$ shell i.e.
$^{96}$Cd , to make a comparison of theory and experiment, but such
a comparison can be made in the f$_{7/2}$ region . The interaction
used for $^{44}$Ti is INTa from the two-particle spectrum of $^{42}$Sc;
for$^{52}$ Fe we use INTb from the two-hole spectrum of $^{54}$Co.

In $^{44}$Ti we have : E(6)-E(4) = 1.561 MeV---E(8)-E(6) =2.493 MeV----
E(10)-E(8) = 1.163 MeV

In $^{52}$Fe we have : E(6)-E(4)= 1.941 MeV --- E(8) -E(6) =2.035
MeV---E(10)- E(8) = 1.021 MeV.

These empirical results are in qualitative agreement with the predictions
of the single j shell model-that there is indeed a midshell gap in
energies of levels below midshell and those above midshell. The effect
is not as pronounced in $^{52}$Fe as it is in $^{44}$Ti but it is
there nevertheless. It would be of great interest to find more details
of the spectra of $^{96}$Cd and $^{96}$Ag to see if indeed there
is such a gap in these heavier nuclei.

. Table VI Comparison of experiment and theory (INT) for the gaps.

\begin{tabular}{|c|c|c|c|}
\hline 
$^{44}$Ti & MeV & EXPT. & INTa\tabularnewline
\hline 
\hline 
INTa & E(6)-E(4) & 1.5611 & 1.272\tabularnewline
\hline 
 & E(8)-E(6) & 2.493 & 2.024\tabularnewline
\hline 
 & E(10)-E(8) & 1.163 & 1.307\tabularnewline
\hline 
$^{52}$Fe &  & EXPT. & INTb\tabularnewline
\hline 
INTb & E(6)-E(4) & 1.941 & 1.403\tabularnewline
\hline 
 & E(8)-E(6) & 2.035 & 1.783\tabularnewline
\hline 
 & E(10)-E(8) & 1.021 & 1.072\tabularnewline
\hline 
\end{tabular}

\section{Explanation via Pairing interaction}

We feel we can explain the above gap in the spectrum via the pairing
interaction of B.H. Flowers {[}8{]} and A.R. Edmonds and B.H. Flowers
{[}9{]}. The two body matrix elements in say the g$_{9/2}$ shell
from J=0 to J=9 are -A,0,0,0,0,0,0,0,0,0, with A positive. The expression
for the energies is :

$E=$C {[} (n-v)/4 {*}(4j+8-n-v) -T(T+1+ t(t+1){]}.

with C negative. Here n is the number of valence nucleons, v is the
seniority, T is the total isospin and t is the reduced isospin. The
relation between A and C has ben discusssed by Harper and Zamick {[}10{]}.
They note that for any j shell C=-A/((2j+1). This can be obtained
by noting that the quantum numbers (T,t,v) for the lowest I=0T=0 state
are (0,0,0) whilst for the unique I=0 T=2 state they are (2,0,0).

In the N=Z nuclei we are dealing only with T=0 states whilst for the
odd-odd nuclei in ref {[}1{]} all states had T=1.
\begin{description}
\item [{The}] resulting yrast spectra for even I are as follows when we
choose A=1 : :E(0)=0 , E(I)=1 for I=2,4,6and 8, E(I)= 2.2 for I=10,
12, 14and 16. So we see there is a break after I=8. We can understand
this because the energy for this interaction does not depend explicitly
on I. But it does depend on seniority. For I=2,4,6,and 8 we can have
seniority v=2 states by coupling one pair of nucleons to J=0. But
we cannot reach I=10,12,14 or 16 by coupling one pair to to J=0. Hence
the latter states must have seniority v=4.
\end{description}
We can also look at the odd specra. For I= 1,3,5,7,9 E(I) is equal
to 1.4 whilst for I= 11, 13 and 15 E(I)=2.2 Hence there is a predicted
gap between I=9 and I=11. The same seniority argument applies for
odd I.  Such a gap also appears with more realistic interactions as
seen in TableV.

It should be emphasized that for most interactions e.g. INT seniority
is not a good quantum number for a system of both neutrons and protons.
With the interactions of Edmonds and Flowers {[}7,8{]} seniority is
a good quantum number . Some remnants of the senioriy behaviour in
their simple interaction seem to have not been completely lost in
the more complex INT and other interactions.

For completenss we show results with shematic interactions for yrast
states in $^{96}$Cd in Table IX and for yrast states in $^{96}$Ag
in Table X.

The iteractions are:

E(0) J=0 pairing -1.0 , 0,0,0,0,0,0,0,0,0

E(9) J$_{max}$pairing 0,0,0,0,0,0,0,0,0, -1.0

E(0,9) Sum of above -1.0, 0,0,0,0,0,0,0,0,-1.0

Note that with E(0) there is a high degeneracy--the low lying spectra
do not spread out. Also note that with E(9) the ground state does
not have I=0. Rather it has I=16, the maximum I , as the grounsd state.
Although both E(0) and E(9) yield terrible spectra when we mix them
to form E(0,9) we have the beginning of a reasonable spectrum, with
I=0 as the ground state and some spreading out of the energy levels.

\section{Appendix: Interactions used}

The interactions used in the f$_{7/2}$ shell are shown in Table VII.
The interactions used in the g$_{9/2}$ shell are shown in Table VIII.
We also show the Q.Q interaction.In some but not all cases a constant
has been added so that the J=0 matrix element is zero. This does not
affect the spectra.

Table VII: f$_{7/2}$matrix elements 

\begin{tabular}{|c|c|c|c|}
\hline 
J & $^{42}$Sc & $^{54}$Co & Q.Q\tabularnewline
\hline 
\hline 
0 & 0 & 0 & 0\tabularnewline
\hline 
1 & 0.6111 & 0.5723 & 0.4096\tabularnewline
\hline 
2 & 1.5863 & 1.4465 & 1.1471\tabularnewline
\hline 
3 & 1.4904 & 1.8244 & 2.0483\tabularnewline
\hline 
4 & 2.8153 & 2.6450 & 2.8677\tabularnewline
\hline 
5 & 1.5101 & 2.1490 & 3.2774\tabularnewline
\hline 
6 & 3.2420 & 2.9600 & 2.8677\tabularnewline
\hline 
7 & 0.6163 & 0.1990 & 1.1471\tabularnewline
\hline 
\end{tabular}

.

.Table VIII g$_{9/2}$matrix elements

\begin{tabular}{|c|c|c|c|c|}
\hline 
J & INTd & Qi & Corragio & Q.Q\tabularnewline
\hline 
\hline 
0 & 0.0000 & 0.0000 & -2.3170 & -1.0000\tabularnewline
\hline 
1 & 1.1387 & 1.2200 & -1.4880 & -0.8788\tabularnewline
\hline 
2 & 1.3947 & 1.4580 & -0.6670 & -0.6515\tabularnewline
\hline 
3 & 1.8230 & 1.5920 & -0.4400 & -0.3485\tabularnewline
\hline 
4 & 2.0283 & 2.2830 & -0.1000 & -0.0152\tabularnewline
\hline 
5 & 1.9215 & 1.8820 & -0.2710 & 0.2789\tabularnewline
\hline 
6 & 2.2802 & 2.5490 & 0.0660 & 0.4848\tabularnewline
\hline 
7 & 1.8797 & 1.9300 & -0.4040 & 0.4848\tabularnewline
\hline 
8 & 2.4275 & 2.6880 & 0.2100 & 0.1818\tabularnewline
\hline 
9 & 0.7500 & 0.6260 & -1.4020 & -0.5454\tabularnewline
\hline 
\end{tabular}

.

.Table IX Calculated spectra with schematic interactions $^{96}$Cd

.%
\begin{tabular}{|c|c|c|c|}
\hline 
I & E(0) & E(9) & E(0,9)\tabularnewline
\hline 
\hline 
EVEN &  &  & \tabularnewline
\hline 
0 & 0.0000 & 0.5294 & 0.0000\tabularnewline
\hline 
2 & 1 & 0.5294 & 0.6370\tabularnewline
\hline 
4 & 1 & 0.5294 & 0.9292\tabularnewline
\hline 
6 & 1 & 0.5294 & 1.1965\tabularnewline
\hline 
8 & 1 & 0.5286 & 1.2562\tabularnewline
\hline 
10 & 2.2 & 0.5253 & 1.6071\tabularnewline
\hline 
12 & 2.2 & 0.4835 & 1.5674\tabularnewline
\hline 
14 & 2.2 & 0.3285 & 1.4124\tabularnewline
\hline 
16 & 2.2 & 0.0000 & 1.0839\tabularnewline
\hline 
ODD &  &  & \tabularnewline
\hline 
1 & 1.4 & 1.5298 & 2.4323\tabularnewline
\hline 
3 & 1.4 & 1.5278 & 2.1862\tabularnewline
\hline 
5 & 1.4 & 1.5114 & 2.1059\tabularnewline
\hline 
7 & 1.4 & 1.4247 & 2.1113\tabularnewline
\hline 
9 & 1.4 & 1.0293 & 1.4943\tabularnewline
\hline 
11 & 2.2 & 1.0247 & 2.1086\tabularnewline
\hline 
13 & 2.2 & 0.9700 & 2.0539\tabularnewline
\hline 
15 & 2.2 & 0.7941 & 1.8787\tabularnewline
\hline 
\end{tabular}

.

.Table X Calculated Spectra with schematic interactions $^{96}$Ag

\begin{tabular}{|c|c|c|c|}
\hline 
I & E(0) & E(9) & E(0,9)\tabularnewline
\hline 
\hline 
0 & 0.0000 & 1.735 & 0.5190\tabularnewline
\hline 
1 & 0.8 & 0.7357 & 09380\tabularnewline
\hline 
2 & 0.6 & 0.7349 & 0.5310\tabularnewline
\hline 
3 & 0.8 & 0.7337 & 0.6919\tabularnewline
\hline 
4 & 0.6 & 0.7213 & 0.3025\tabularnewline
\hline 
5 & 0.8 & 0.7173 & 0.6115\tabularnewline
\hline 
6 & 0.6 & 0.6325 & 0.2361\tabularnewline
\hline 
7 & 0.8 & 0.6306 & 0.6170\tabularnewline
\hline 
8 & 0.6 & 0.2353 & 0.0544\tabularnewline
\hline 
9 & 0.8 & 0.2352 & 0.0000\tabularnewline
\hline 
10 & 1.6 & 0.2360 & 0.6196\tabularnewline
\hline 
11 & 1.6 & 0.2306 & 0.6142\tabularnewline
\hline 
12 & 1.6 & 0.2504 & 0.6340\tabularnewline
\hline 
13 & 1.6 & 0.1759 & 0.5596\tabularnewline
\hline 
14 & 1.6 & 0.4412 & 0.8248\tabularnewline
\hline 
15 & 1.6 & 0.0000 & 0.3837\tabularnewline
\hline 
\end{tabular}

\end{document}